\documentclass[aps,prl,twocolumn,showpacs]{revtex4}
\usepackage{stmaryrd}
\usepackage{amssymb}

\usepackage{pifont}
\usepackage{txfonts}
\usepackage{graphicx}

\begin{document}

\title{Synthesizing and characterization of hole doped nickel based layer superconductor (La$_{1-x}$Sr$_{x}$)ONiAs}

\author{Lei Fang, Huan Yang, Peng Cheng, Xiyu Zhu, Gang Mu }
\author{Hai-Hu Wen}\email{hhwen@aphy.iphy.ac.cn}

\affiliation{National Laboratory for Superconductivity, Institute of
Physics and Beijing National Laboratory for Condensed Matter
Physics, Chinese Academy of Sciences, P.O. Box 603, Beijing 100190,
People's Republic of China}

\begin{abstract}
We report the synthesizing and characterization of the hole doped
Ni-based superconductor ($La_{1-x}Sr_{x})ONiAs$. By substituting La
with Sr, the superconducting transition temperature $T_c$ is
increased from 2.75 K of the parent phase $LaONiAs$ to 3.7 K at the
doping levels x= 0.1 - 0.2. The curve $T_c$ versus hole
concentration shows a symmetric behavior as the electron doped
samples $La(O_{1-x}F_{x})NiAs$. The normal state resistivity in
Ni-based samples shows a good metallic behavior and reveals the
absence of an anomaly which appears in the Fe-based system at about
150 K, suggesting that this anomaly is not a common feature for all
systems. Hall effect measurements indicate that the electron
conduction in the parent phase $LaONiAs$ is dominated by
electron-like charge carriers, while with more Sr doping, a
hole-like band will emerge and finally prevail over the conduction,
and accordingly the superconducting transition temperature $T_c$
increases.

\end{abstract}

\maketitle

 Searching for high temperature superconductors has been a
 long-term strategy in material science. Superconductors
 with unconventional pairing symmetry
 found in past decades seem to
 have some common features: layered structure, such as in
 cuprates\cite{1}; tunable transition
 temperature (T$_{c}$) by doping holes or electrons; possible exotic pairing mechanism rather than phonon mediated
 superconductivity, for instance in the heavy fermion system\cite{2}.  The newly discovered
 iron-based
 superconductor La(O$_{1-x}$F$_{x}$)FeAs with a moderate high T$_c$ = 26 K
 seems to fit to these three categories\cite{3,4,MuG,ShanL}. It is found that La(O$_{1-x}$F$_{x}$)FeAs
 belongs to a layered structure constructed by stacking the LaO and FeAs sheets alternatively, where FeAs sheet is regarded as the conduction layer
 whose charge carrier density could be tuned by the neighboring LaO
 sheet by charge doping. Substituting part of the oxygen with fluorine, the system
 changes from having a weak insulating behavior to superconductive with x = 0.05 - 0.12\cite{4}. This discovery
 has stimulated intense efforts in both experimental and theoretical
studies.  Theoretically it was concluded that the correlation of
this system could be moderate\cite{5,6}. Experimentally both low
temperature specific heat\cite{MuG} and point contact
tunneling\cite{ShanL} measurements indicate the possible
unconventional pairing symmetry. Very recently, superconductivity at
about 25 K was found in hole doped samples ($La_{1-x}Sr_x)OFeAs$ by
our group\cite{7}. In present work we report the fabrication and
characterization of the hole doped Ni-based superconductors
(La$_{1-x}$Sr$_{x}$)ONiAs. Superconductivity at about 3.7 K was
found and the $T_c$ exhibits a symmetric behavior in both hole-doped
and electron-doped side. The hole-like charge
     carriers in the present Sr doped samples are evidenced by Hall effect measurements.

Polycrystalline  (La$_{1-x}$Sr$_{x}$)ONiAs samples ( x = 0.1, 0.2,
0.3) were synthesized by the conventional solid state reaction
method. Stoichiometric LaAs powder was home made by reacting pure
$La$ (99.99\%) and $As$ (99.99\%). Later it is mixed with dehydrated
La$_{2}$O$_{3}$(99.9\%), SrO (99.5\%), and NiAs powder (home made by
reacting pure $Ni$ (99.99\%) and $As$ (99.99\%)),
     and Ni powder (99.99\%), grounded and pressed into a
     pellet. Then the pellet was sealed into an evacuated quartz tube.
     Consequently, the tube was slowly warmed up in a muffle furnace to 1150 $^{\circ}$ C
     and sintered for 48 hours, then cooled down to room
     temperature. X-ray diffraction (XRD) pattern measurement was performed at room temperature employing an M18AHF
     x-ray diffractometer (MAC Science).
     The magnetic measurements
     were carried out on a Magnetic Property Measurement System (MPMS, Quantum Design). The electrical resistivity and Hall coefficient
     were measured by a six-probe method based on a Physical Property Measurement System (PPMS, Quantum Design).

\begin{figure}

       \includegraphics[width=8cm]{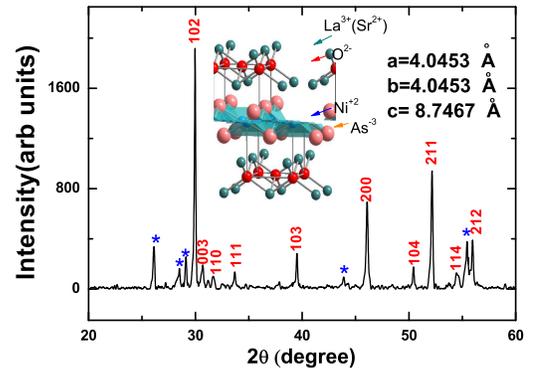}
       \caption{XRD pattern of (La$_{1-x}$Sr$_{x}$)ONiAs with x = 0.2, which can be indexed in a
     tetragonal symmetry with a = b = 4.0453 ${\AA}$ and c = 8.7467 ${\AA}$. The asterisks mark the peaks from impurity phase.
     }
       \label{figure1}
\end{figure}

     Fig.1 (a) shows the XRD pattern of the
     sample (La$_{0.8}$Sr$_{0.2}$)ONiAs, which can be indexed in a
     tetragonal space group with \emph{a} = \emph{b} = 4.0453 ${\AA}$ and \emph{c} = 8.7467 ${\AA}$.
     Though minor peaks arising from the impurity phase were found (could come from NiAs),
     there is no doubt that the main phase is dominated by (La$_{1-x}$Sr$_{x}$)ONiAs in the sample with x = 0.20 and below. It should
     be noted that the indices of the present sample are a bit away from
     that of LaONiAs\cite{8}( a = b = 4.119 ${\AA}$,
     c = 8.18 ${\AA}$) and close to that of (La$_{0.9}$Sr$_{0.1}$)OFeAs\cite{7}, especially along the \emph{c
     } direction. A possible explanation is that the lattice of $LaONiAs$ is
     distorted by an incommensurate replacement of La$^{3+}$ by Sr$^{2+}$.
Taking account of the layered structure of $LaONiAs$, the cell
     lattice could be inclined to compromise the divalent strontium
     with a bigger ironic radius(1.12 ${\AA}$). It is observed that the crystalline quality of
     (La$_{0.8}$Sr$_{0.2}$)ONiAs and (La$_{0.9}$Sr$_{0.1}$)ONiAs are
     similar, whereas a lot of impurity peaks appear in
     the sample with x = 0.3, indicating stronger segregation during the sintering. This imperfect chemical reaction may
     be improved by taking a reaction at a higher temperature.
     But this is difficult to do with the quartz tube since it becomes softened at a temperature of
     about 1180 $^{\circ}$ C. Therefore in the following discussion,
     the data of (La$_{0.7}$Sr$_{0.3}$)ONiAs are not included.

\begin{figure}

       \includegraphics[width=9cm]{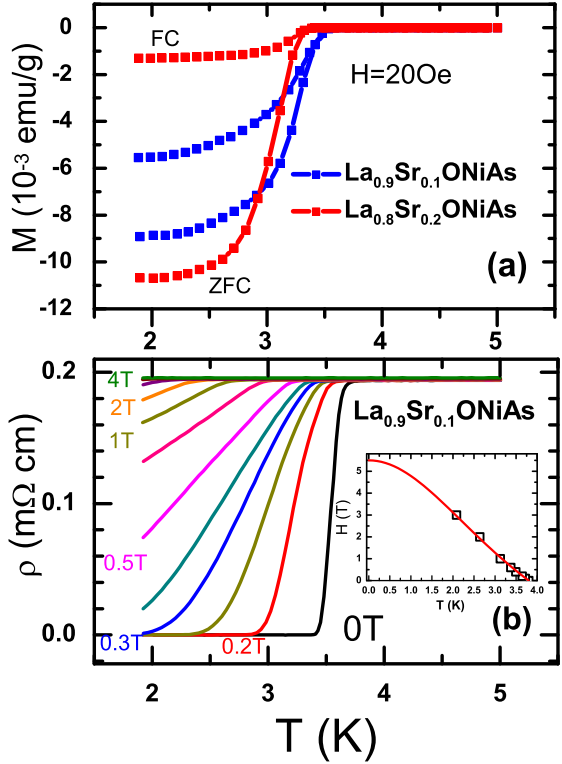}
       \caption{(a) DC magnetization of (La$_{1-x}$Sr$_{x}$)ONiAs samples with x = 0.1 and 0.2, measured in the zero-field-cooled
       (ZFC) and field-cooed (FC) processes. The superconducting fraction estimated at 2 K is beyond 40 \%. (b) The temperature dependence of resistivity of the sample
       with x = 0.1 under different magnetic fields. It is clear that the superconducting transition is broadened by using a magnetic field. The upper critical field is
       determined with the criterion $\rho=95\rho_n$ and shown as an inset of Fig.2(b). The solid line in the inset shows the theoretical curve based on the GL theory (see text).}
       \label{figure2}
\end{figure}

 The DC magnetization data of (La$_{1-x}$Sr$_{x}$)ONiAs were shown
 in Fig. 2(a). A simple estimation on the magnetization at 2K tells that the superconducting volume fraction is beyond 40\% for the sample with x = 0.2.
  Fig.2(b) shows the temperature dependence of resistivity under different magnetic fields. A sharp transition with the width of about 0.4 K is observed at 3.7 K.
  By applying a magnetic field, the resistive transition curve broadens quickly showing a strong vortex flow behavior. But the onset transition point, which is close to the upper critical field, moves slowly with the magnetic
  field. This is similar to that observed in F-doped
  $LaOFeAs$\cite{ZhuXY}. Compared with the pure phase LaONiAs with T$_c\approx$ 2.75 K\cite{8},  T$_c$
 of strontium substituted samples are improved to 3.7 K and 3.5 K for doping x=0.1 and 0.2,
 respectively. According to
   Ginzburg-Landau theory, zero temperature upper critical field
   H$_{c2}(0)$
   could be derived from the formula:
   H$_{c2}(T)$=H$_{c2}(0)$(1-t$^{2}$)/(1+t$^{2}$), where t is
   the normalized temperature T/T$_{c}$. It is found that the theoretical curve can describe the experimental data very well. The derived H$_{c2}(0)$ is
   found to be
   about 5.5 T, being close to that in the F-doped Ni-based system\cite{8}.

   Fig.3(a) shows the resistivity of (La$_{1-x}$Sr$_{x}$)ONiAs
   with x = 0.1, 0.2 from 2 K to 300 K at zero field. The resistivity in the normal state for all doping levels show metallic
   behavior. Near 3.7 K the resistivity of (La$_{0.9}$Sr$_{0.1}$)ONiAs
   drops sharply to zero, whereas, the
   risistivity of
   (La$_{0.8}$Sr$_{0.2}$)ONiAs drops at about 3.5 K with a similar transition width. For a better comparison, the resistivity of
   La(O$_{0.9}$F$_{0.1}$)NiAs with (T$_{c} \approx $3.8K) was also shown in Fig.
   3(a). It is interesting to note that, at all doping levels the normal state resistivity
   of the present Ni-based system
   exhibit no anomaly as found in the F-doped Fe-based system at
   about 150 K. Recently it was argued that this anomaly could be due to the formation of a competing order, such as spin density wave
or charge density wave\cite{SDW}. We would argue that this
resistivity anomaly observed in the Fe-based system may not be a
common feature for all systems, therefore it is too early to say
whether there is a close relationship between this anomaly and
superconductivity or not.

In Fig.3(b) we show an enlarged view for the resistive transitions
for samples with x=0.1 and 0.2. The transition temperature of sample
x = 0.2 is about 3.5 K, which is very close to that of sample x=0.1,
but obviously higher than that of the undoped parent phase $LaONiAs$
$T_c \approx 2.75 K$. Interestingly, if we plot the $T_c$ versus the
hole concentration, the curve exhibits a symmetric behavior with the
electron doped side\cite{8}. This behavior has also been found in
our original work for hole doped  $(La_{1-x}Sr_x)OFeAs$ system. The
similar behavior in both systems may suggest that the density of
states in the two sides of the Fermi energy is roughly symmetric,
which has actually been claimed already by the calculations based on
the dynamical mean field theory (DMFT)\cite{6}.

\begin{figure}

       \includegraphics[width=7cm]{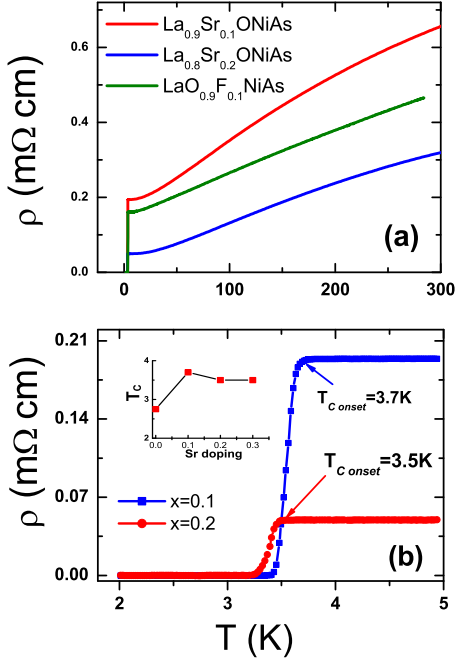}
       \caption{(a)Temperature dependence of  resistivity in wide temperature region for samples with x=0.1 and 0.2, and the electron doped sample La(O$_{0.9}$F$_{0.1}$)NiAs.
       The normal state does not exhibit an anomaly which appears in the Fe-based system. (b) An enlarged view for the resistive transitions of the samples
       with x=0.1 and 0.2. The inset in Fig.3(b) presents the hole doping dependence of the transition temperature. Combining the data from the electron doped
       side, it is found that the curve of $T_c$ vs. hole and electron
       concentrations exhibits a symmetric behavior.}
       \label{figure3}
\end{figure}

\begin{figure}

       \includegraphics[width=7cm]{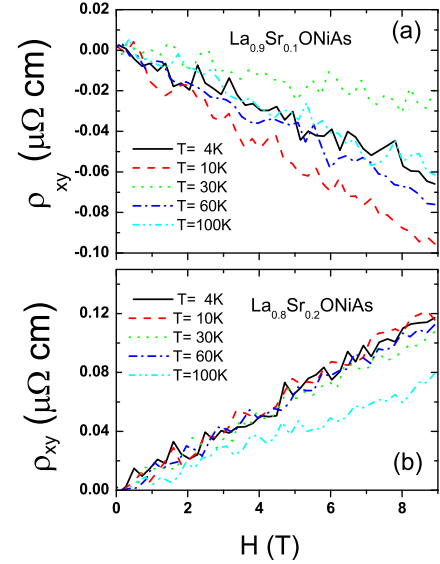}
       \caption{Hall resistivity as a function of applied magnetic field for samples (La$_{1-x}$Sr$_{x}$)ONiAs, x = 0.1 (a) and 0.2 (b),
       respectively. The Hall resistivity is small in magnitude
       compared with the electron doped or undoped samples, indicating the gradual emergence of a hole-like conduction band.}
       \label{figure4}
\end{figure}

Since part of La$^{3+}$ are substituted by Sr$^{2+}$, hole typed
carriers are expected in our present Sr-doped system. A prove to
that by Hall effect measurements is necessary. Fig. 4(a) and Fig.
4(b) show the Hall resistivity $\rho_{xy}$ for sample x = 0.1 and
0.2, respectively. Interestingly, the sign of  $\rho_{xy}$ for x =
0.1 is still negative, but quite close to zero. This is reasonable
since the parent phase $LaONiAs$ is actually dominated by an
electron-like band\cite{8}, the Hall coefficient defined as
$R_H=\rho_{xy}/H$ is -5$\times10^{-10}m^{3}/C$ at 100 K for the
undoped sample. This means that holes are really introduced into the
system by doping Sr. By doping more Sr into the system, the Hall
resistivity $\rho_{xy}$ becomes positive and hole-like charge
carriers finally dominate the conduction at the doping level x =
0.2. Fig.5 presents the Hall coefficient  for two samples below 100
K. It is clear that (La$_{0.9}$Sr$_{0.1}$)ONiAs has more
electron-like charge carriers, but the sample
(La$_{0.8}$Sr$_{0.2}$)ONiAs shows clearly the dominant conduction by
hole-like charge carriers. Our data suggest that with the
substitution of La$^{3+}$ by Sr$^{2+}$, the conduction by the
electron-like band which appears for the undoped phase will be
prevailed over by the hole-like band, and superconductivity at about
3.5-3.8 K occurs when the hole-like band dominates the conduction.

\begin{figure}

       \includegraphics[width=7cm]{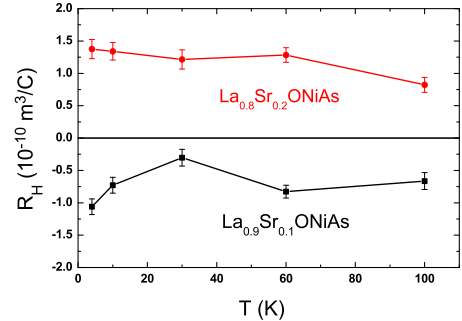}
       \caption{Hall coefficients for samples (La$_{1-x}$Sr$_{x}$)ONiAs with x = 0.1 and 0.2. A sign change is obvious
       with increasing Sr content from 0.1 to 0.2 indicating a dominant conduction by hole-like charge carriers at x = 0.2.}
       \label{figure5}
\end{figure}

In summary, by substituting La with Sr in $LaONiAs$, a systematic
change of both the superconducting transition temperature and normal
state Hall coefficient are observed. First the transition
temperature is increased from 2.75 K to about 3.5 - 3.8 K with Sr
doping, meanwhile the Hall coefficient changes from negative to
positive. The curve of $T_c$ vs. the hole concentration exhibits a
symmetric behavior as the electron doped side, which may suggest a
roughly symmetric distribution of DOS above and below the Fermi
energy. Our data further support the conclusion that
superconductivity can be induced by hole doping.

\begin{acknowledgments}
We acknowledge the fruitful discussions with Yupeng Wang, Zidan Wang
and Tao Xiang. This work was financially supported by the NSF of
China, the MOST of China (973 Projects No. 2006CB601000, No.
2006CB921802 and 2006CB921300), and CAS (Project ITSNEM).
\end{acknowledgments}

\end{document}